\documentclass[12pt,thmsa]{article}
\usepackage{sw20lart}

\input tcilatex
\QQQ{Language}{
American English
}

\begin{document}

\title{\textbf{The Quantum Adiabatic Approximation in Chaotic Systems}}
\author{G. Abal\thanks{%
corresponding author: Gonzalo Abal, Instituto de F\'{\i}sica, Facultad de
Ingenier\'{\i}a, C.C. 30, C.P. 11000, Montevideo, Uruguay. Phone: (5982)
710905/715444/715445. Fax: (5982)\ 711630. email:\ abal@fing.edu.uy}, A.J.
Pereira, A. Romanelli and A. Sicardi-Schifino \\
Instituto de F\'{\i}sica\\
Facultad de Ingenier\'{\i}a\\
C.C. 30, C.P. 11000, Montevideo, Uruguay\\
and\\
Instituto de F\'{\i}sica\\
Facultad de Ciencias Exactas y Naturales\\
C.C. 10773, Montevideo, Uruguay}
\maketitle

\begin{abstract}
\textit{Quantum systems with chaotic classical counterparts cannot be
treated by perturbative techniques or any kind of adiabatic approximations.
This is so, in spite of the quantum suppression of classical chaos. We
explicitly calculate the adiabaticity parameter for the case of a Fermi
Accelerator and show that the adiabatic expansion of the evolution operator
is divergent. The relevance of this situation for the understanding of
quantal effects in the one body dissipation mechanism in nuclear reactions
is briefly discussed. }

\begin{description}
\item  PACS numbers:\ 05.45.+b, 03.65.-w, 24.60.Lz, 72.15.Rn

\item  keywords: adiabatic approximation, dissipative systems, quantum
chaos, quantum localization, one body dissipation.
\end{description}
\end{abstract}

Over the last years, considerable effort has been directed towards the study
of the quantum evolution of simple (few degrees of freedom) classically
chaotic systems. Examples of these kind of systems are the well-known Kicked
Rotor \cite{Casati1} and the Fermi Accelerator \cite{Jose}-\cite{Dembinski}.
This last system is of particular interest since, at the classical level, it
is a model for the One Body Dissipation mechanism \cite{Blocki}. Its
relevance for the description of nuclear reactions at intermediate energies
is now being investigated, as new knowledge on the quantum suppression of
classical chaos becomes available \cite{One Body}. There is increasing
evidence \cite{Evidence} that if quantum effects are taken into account,
they will tend to inhibit this mechanism. The adiabatic approximation has
been commonly used in nuclear physics whenever different time scales are
present. The issue of wether the use of this approximation in nuclei is
actually justified has been previously examined by Nazarewicz \cite
{Nazarewicz}. We address this question in the light of recent developments
in the understanding of the quantum description of classically chaotic
systems. In spite of the quantum suppression of classical chaos, some
characteristics of chaotic systems are still present at the quantum level 
\cite{Haake}. It is therefore natural to ask wether the adiabatic
approximation can be used successfully for quantum systems that are
classically chaotic. It is well established, since the pioneering work by H.
Poincar\'{e} \cite{Poincare} at the end of the last century, that the
adiabatic approximation does not hold in the case of classical chaotic
systems because the perturbative adiabatic series expansions diverge. In
this letter, it is argued that perturbative adiabatic approaches are also
divergent for quantum systems with chaotic classical analogs. We explicitly
work out the case of a quantum Fermi Accelerator model and discuss other
similar examples which cannot be treated by adiabatic techniques. Finally, 
\textit{\ }the relevance of this situation for the understanding of quantum
mechanical effects in the one body dissipation mechanism in nuclear
reactions is briefly discussed.

The Fermi Accelerator \cite{L+L} consists of a particle confined in an
infinite well with one periodically moving wall. The motion of the wall can
be parametrized as $L(t)=L_0\left[ 1+\delta f(t)\right] $, where $\delta \,$
is the dimensionless amplitude of the wall motion and $f(t)$ is a periodic
function scaled so that $|f(t)|\leq 1$. The adiabatic approximation
corresponds to the limit $\delta \rightarrow 0.$ In the position
representation the Hamiltonian operator is $\mathbf{H}=-\frac{\hbar ^2}{2m}%
\frac{\partial ^2}{\partial x^2},$ with time dependent boundary conditions
imposed on the wavefunction $\Psi (x=0,t)=\Psi (x=L(t),t)=0.$ No exact
solution is known for this problem unless $\delta =0$. Following Makowski 
\cite{Makowski}, we use the transformation $y=\frac xL$ and after an
adequate redefinition of the phase of the wavefunction, the problem can be
stated in terms of time independent boundary conditions $\varphi
(y=0,t)=\varphi (y=1,t)=0.$ This can be done at the cost of introducing a
time dependent potential in the transformed Hamiltonian operator 
\begin{equation}
\widetilde{\mathbf{H}}=-\frac{\hbar ^2}{2mL^2}\frac{\partial ^2}{\partial y^2%
}+\frac 12mL\frac{d^2L}{dt^2}y^2.  \label{Ham2}
\end{equation}
For the cases of a sawtooth and triangular wave wall motions, this
Hamiltonian has recently been shown to be equivalent to a generalized Kicked
Rotor Hamiltonian \cite{Zaslavsky}. In these cases, the function $L(t)$ is
not analytic. The evolution operator of impulsive systems, such as the
Kicked Rotor, can be obtained explicitly and a quantum map which describes
the time evolution of the wavefunction can be constructed \cite{Ott}. On the
other hand, as stated above, in the case of the Fermi Accelerator with a
periodically moving wall described by an analytic function of time, the
evolution operator has not been obtained in closed form. In fact, the task
of obtaining a closed expression for the evolution operator for the
important case of sinusoidal wall motion has proved to be a very elusive one 
\cite{Makowski},\cite{Seba}.

The fact that the effective Hamiltonian for this problem, $\widetilde{%
\mathbf{H}},$ can be written (Eq. \ref{Ham2}) as the sum of an integrable
term plus a time dependent potential which is proportional to the
perturbation strength $\delta ,$ has motivated certain authors\cite{Hussein},%
\cite{Pereshogin} to suggest the use of perturbative techniques to find an
approximate solution for the Schr\"{o}dinger equation valid in the adiabatic
limit $\left( \delta \rightarrow 0\right) .$ The classical Fermi Accelerator
is strongly chaotic in the low velocity limit \cite{L+L} and thus it is an
example of a non-adiabatic system. This is a consequence of the topological
changes that the phase space suffers in going from an integrable problem $%
(\delta =0)$ to a chaotic problem $\left( \delta \neq 0\right) .$ At the
quantum level, the quantum suppression of chaos leaves the question open as
wether the adiabatic approximation can be used or not. In the Kicked Rotor,
the quantum suppression of classical chaos has been well understood \cite
{Grempel} in terms of a mathematical analogy with Anderson localization in
solid state physics \cite{Anderson}. Recent numerical results \cite{Abal}\
show that a similar suppression of classical chaos is present in the quantum
Fermi Accelerator with sinusoidal wall motion and suggest that strong
similarities exist between both systems as other authors \cite{Casati2},\cite
{Zaslavsky} have pointed out previously. Therefore, the question of wether
the adiabatic approximation can be used in the quantum version of these
systems, in spite of their classically chaotic nature, does not have an
obvious answer.

The general formalism for an adiabatic expansion of the evolution operator
of a time-dependent quantum Hamiltonian has been developed recently by A.
Mostafazadeh\cite{Mustafa}. There, a series expansion of the evolution
operator in orders of $\nu ,$ the adiabaticity parameter defined below, is
presented 
\begin{equation}
\mathbf{U}(t)=\mathbf{U}^{(0)}(t)\left[ \mathbf{T}e^{-\frac i\hbar \int_0^tds%
\mathbf{H}^{\prime }(s)}\right] =\sum_{p=0}^{N-1}\mathbf{U}^{(p)}(t)+%
\mathcal{O}(\nu ^N)  \label{evolution}
\end{equation}
where $\mathbf{U}^{(0)}$ is the evolution operator in the absence of a
perturbation $(\delta =0)$, $\mathbf{T}$ is the time ordering operator and $%
\mathbf{H}^{\prime }(t)$ is a transformed Hamiltonian defined in \cite
{Mustafa}. The matrix elements of $\mathbf{H}^{\prime }$ are proportional
(in absolute value)\ to those of the matrix $\mathbf{A}$ defined below. We
show that this kind of expansion of the evolution operator does not converge
for the Fermi Accelerator problem with a periodic wall motion. This is so,
because the adiabaticity parameter $\nu $ is of the order of the dimension
of the Hilbert space of the problem and is therefore unbounded.

Let us calculate naively the adiabaticity parameter defined in \cite{Mustafa}
for the Hamiltonian $\mathbf{H}$ of the Fermi Accelerator problem defined
above. We introduce the dimensionless time $\tau =\omega t=\frac{2\pi t}T$ ,
in terms of the period of the forcing function $f(t).$ The instantaneous
eigenstates satisfy $\mathbf{H}|n,\tau >=\varepsilon _n(\tau )|\,n,\tau >$
with $\varepsilon _n(\tau )=\varepsilon _0\frac{L_0^2}{L^2(\tau )}.$ The
adiabaticity parameter is defined as $\nu \equiv \frac 1{\varepsilon
_0}Sup\left[ \mathbf{A}_{mn}\right] ,$ where $\varepsilon _0\equiv \frac{%
\hbar \pi ^2}{2m\omega L_0^2}$ is a convenient energy scale and the matrix $%
\mathbf{A}_{mn}(\tau ),$ related to the rate of change of the instantaneous
eigenstates, is defined (for $m\neq n)$ \textbf{\ }as 
\begin{equation}
\mathbf{A}_{mn}\equiv <m,\tau \,|\frac d{d\tau }|\,n,\tau >.  \label{Amn}
\end{equation}
We denote by $Sup[\mathbf{A}_{mn}]$ the minimum upper bound of the elements
of the matrix $\mathbf{A}$. In the position representation, the
instantaneous eigenstates are $<x,\tau \,|\,n,\tau >=\sqrt{\frac 2{L(\tau )}}%
\sin \left( \frac{n\pi x}{L(\tau )}\right) .$ The explicit expression for
the adiabaticity parameter is 
\begin{equation}
\nu =\frac 2{\varepsilon _0}Sup\left[ \frac 1L\left| \frac{dL}{d\tau }\frac{%
mn}{m^2-n^2}\right| \right] .  \label{nu}
\end{equation}
It is clear from the above equation that for $L(\tau )$ periodic and $\delta
\neq 0$, the adiabaticity parameter diverges as $m,n\rightarrow \infty $.
The series expansion for $\mathbf{U}$ (Eq. \ref{evolution}) does not
converge in this case and the divergence cannot be resolved by a
redefinition of the adiabaticity parameter \cite{Mustafa}. A similar
analysis shows that the same is true for impulsive systems such as the
Kicked Rotor, the Quantum Bouncer \cite{Dembinski} and other related quantum
systems with classical chaotic behavior.

We conclude that periodically driven quantum systems with chaotic classical
analogs, cannot be properly treated by perturbative techniques or adiabatic
approximations since in these cases, the adiabaticity parameter is unbounded
and the series expansion of $\mathbf{U}(t)$ is divergent. Thus, in spite of
the quantum suppression of classical chaos, the non-adiabatic nature of
classically chaotic systems subsists at the quantum level. In contrast,
there are related problems which are classically integrable and for which
the adiabatic approximation has been successfully applied \cite{Integrables},%
\cite{Hussein}. In these cases, the moving wall has a final constant
velocity. As a result, the adiabaticity parameter satisfies $\nu <1$ and the
adiabatic expansion of $\mathbf{U}$ is convergent. However, integrable
Hamiltonians cannot describe dissipative mechanisms. In particular, in order
to assess the importance of quantum effects on the one body dissipation
mechanism in nuclei, classically chaotic models must be studied from a
quantum mechanical perspective. The transition from the integrable case ($%
\delta =0)$ and the chaotic case $\left( \delta \neq 0\right) $ introduces a
qualitative change in the nature of the problem, which makes it essentially
non-adiabatic and as a consequence it cannot be treated successfully by any
form of adiabatic perturbation theory.

\bigskip\ 

We aknowledge support from PEDECIBA and CONICYT.

\end{document}